\documentclass[A4paper,pre,nofootinbib,twocolumn,showpacs,aps,floats,floatfix,superscriptaddress]{revtex4-1}
\usepackage{enumerate}
\usepackage{multirow}
\usepackage{graphicx}
\usepackage{caption}
\usepackage{subcaption}
\usepackage{color}
\usepackage{graphics}
\usepackage{epsfig}
\usepackage{amsmath,amssymb,bm}
\usepackage{dcolumn}
\usepackage{rotating}
\usepackage{float}
\usepackage{setspace}
\usepackage{footmisc}

\begin{document}

\title{Network communities within and across borders}
\author{Federica Cerina} \affiliation{Department of Physics,
  Universit\`{a} degli Studi di Cagliari, Cagliari, Italy}\affiliation{Linkalab, Complex
  Systems Computational Laboratory, Cagliari 09129, Italy} 
\author{Alessandro Chessa} \affiliation{Linkalab, Complex
  Systems Computational Laboratory, Cagliari 09129, Italy}  \affiliation{IMT Institute 
  for Advanced Studies Lucca, Piazza S. Ponziano 6, 55100 Lucca, Italy}     
\author{Fabio Pammolli}\affiliation{IMT Institute 
  for Advanced Studies Lucca, Piazza S. Ponziano 6, 55100 Lucca, Italy} 
\author{Massimo Riccaboni} \affiliation{IMT Institute 
  for Advanced Studies Lucca, Piazza S. Ponziano 6, 55100 Lucca, Italy}

\begin{abstract}
We investigate the impact of borders on the topology of spatially embedded networks. Indeed territorial subdivisions and geographical borders significantly hamper the geographical span of networks thus playing a key role in the formation of network communities. This is especially important in scientific and technological policy-making, highlighting the interplay between  pressure for the internationalization to lead towards a global innovation system and the administrative borders imposed by the national and regional institutions.
In this study we introduce an outreach index to quantify the impact of borders on the community structure and apply it to the case of the European and US patent co-inventors networks.
We find that (a) the US connectivity decays as a power of distance, whereas we observe a faster exponential decay for Europe; (b) European network communities essentially correspond to nations and contiguous regions while US communities span multiple states across the whole country without any characteristic geographic scale. We confirm our findings by means of a set of simulations aimed at exploring the relationship between different patterns of cross-border community structures and the outreach index. 
\end{abstract}

\maketitle 
\section*{Introduction} \label{sec:intro}

Over recent decades, political change and new transportation and information technologies have enhanced international openness and cross-border integration.
Globalization has made social networks more international and human communities more integrated across cultural and political borders. This is witnessed 
by the increasing number of long-range connections in multiple networks, such as trade, human mobility, communications, financial investments and 
scientific collaborations \cite{arunachalam2000,scherngell2012,hoekman2013,thiemann2010}.

Enabled by modern technology, people from all over the world are offered a myriad of opportunities for social interactions and group assembly with increasingly 
larger geographic ranges \cite{onnela2011}. Nonetheless, this does not mean that networks can stretch across a borderless world indefinitely: as for climate networks 
one can detect geographical regions with the same climate variability \cite{tsonis2004,daqing2011,berezin2012}. As individual nodes in socio-economic networks occupy a 
given region in space, it is reasonable to assume that geographical proximity also plays a crucial role in social link formation \cite{watts1998}. 
Indeed, a power-law decay in link probability with distance acting as a spatial constraint has been observed \cite{onnela2011,daqing2011,lambiotte2008,goldenberg2009}. 
 In a recent meta-analysis estimating the role of distance in international trade, it has been shown that  $t \propto d^{-\gamma}$, where $t$ is trade, $d$ is the distance and $\gamma  \approx 1$ over more than a century of data \cite{disdier08}.
Arguably, distance is not the only spatial constraint on link formation, since natural and artificial borders also have the power to hamper connectivity. Communication 
and transportation routes, on the contrary, facilitate long-distance interactions \cite{brockmann2013}. Geographical and institutional borders are relevant in all networks where distance matters, such as power 
grid networks, transportation and communication networks as well as collaboration networks. Natural, artificial and administrative borders can substantially reduce the probability 
of link formation by introducing a major physical constraint in terms 
of cost, service, capacity and reliability of global networks. Two of the most widely accepted results in international economics are that trade is impeded by distance and that 
the crossing of national borders also sharply reduces trade. It has been shown, for instance, that national borders are responsible for a fivefold decrease in world trade when 
compared to a borderless world \cite{eaton2002}.  Well-known global networks are transportation and communication networks, such as the airline network and the World Wide Web, 
for which the role of borders has been recently documented \cite{halavais2000,guimera2005}. 
 
Although physical and social propinquity still dominates the organization of human activities, increasing long-range interactions are radically transforming 
the architecture of global networks across cultural, political and geographical borders. Different measures of globalization have been used in the 
literature. Traditionally, international openness has been proxied by the 
share of cross-border links over the total number of connections. More recently, various network-based measures of cross-border integration 
have been introduced \cite{kali2007,arribas2009,duernecker2012}. Similarly, a multinational corporation consists in a group of geographically dispersed 
organizations that include its headquarters and the various national subsidiaries. Such an entity can be conceptualized as an international spatially embedded network 
to develop network measures of firm internationalization \cite{ghoshal1990,rauch2001}. Network-based measures take ``who connects with whom'' into consideration, rather than 
just looking at the degree of openness.
On a different plane, the effective borders between spatially embedded networks only partially overlap with existing administrative borders \cite{thiemann2010}.
To properly measure the extent of the international span of networks and
cross-national communities it is of paramount importance to assess the
effectiveness of policies devoted to international collaboration, such as the ones implemented by the European Union to favor the free movement of people, goods, investments and ideas across European borders.
As part of this effort, the European Research Area has been recently deemed equivalent in terms of research and innovation with respect to the European common market for goods and services. 
In this paper we explore the effect of borders on the European and US co-inventorship networks as a way to assess the progress toward
the effective cross country integration of scientific and technological communities \cite{chessa2013}.

We proceed as follows. In ``Methods" we describe the
methodology we used to define an outreach index, while in
``Data" we analyze the patent dataset by building a geo-coded
network of inventors.
Section ``Results" illustrates our main findings on (a) the distance
distribution of links (b) the geographical distribution of network communities
and the selection of core regions and (c) the outreach and geographical span of
communities. As a robustness check we run a set of simulations to show how the
outreach index can be used to identify different patterns of cross-border
communities. The concluding section outlines our contribution to the analysis
of cross-border spatially embedded network communities. We also highlight the
relevance of our methodology to the proper assessment of the future emergence of an
integrated European Research Area.

\section*{Methods}\label{sec:methods}

\subsection*{Community detection and core regions}

Beyond the local topological features, many networks have
groups of nodes marked by the high density of their internal links with respect to the outgoing links
that connect the groups with each other. This is especially true if the nodes are embedded
in space and subject to geographical constraints that tend to segregate them into
spatial communities. This kind of segregation can be even more pronounced if administrative and
political boundaries are present; a proper method for detecting possible
communities in the network could be a way to assess the role of external
geographical constraints. 

Indeed, if geography has such a strong role in link formation, after performing
a community detection analysis we would expect to find well-defined communities of
spatially connected nodes. However, it has been already shown that geographical clusters and network communities 
do not perfectly overlap \cite{thiemann2010}.
There are now many community detection algorithms \cite{Fortunato:2010} and in the following passage we will use modularity optimization
introduced by Newman and Girvan \cite{Newman:2004} through the ``Louvain''
algorithm \cite{Blondel:2008}.


By definition, if the modularity associated with a network has been optimized, every perturbation in the
partition leads to a negative variation in the modularity ($dQ$). If we move a node from a partition
we have $M-1$ possible choices (with M as the number of communities) as possible
targets for the node's new host community. It is possible to define 
the $dQ$ associated with each node as the smallest variation in absolute value (or the closest to 0 since $dQ$ is always 
a negative number) for all the possible choices. This is a measure of how central that node is to its community \cite{deleo:2013}.

In the next section we introduce two measures of geographical span and community
outreach to properly assess the role of borders in network formation.

\subsection*{Geographical span and community outreach}

The geographical dispersion of a community \emph{s}, or \emph{geographical span} $D_s$, can be measured as \cite{onnela2011}:
\begin{equation}
D_s= \frac{1}{n_s} \sum_{i\in \mathit{C}_s} \sqrt{(X_s-x_i)^2+(Y_s-y_i)^2} 
\end{equation}

\noindent where $n_s$ is the number of nodes in the community $C_s$ and $(X_s,Y_s)$ are the coordinates of the geographical center of the community, 
with $X_s=(1/n_s)\sum_{i\in \mathit{C}_s}x_i$ and $Y_s=(1/n_s) \sum_{i\in \mathit{C}_s}y_i$.

\begin{figure}[ht]
 \begin{center}
  \includegraphics[width=2.5in]{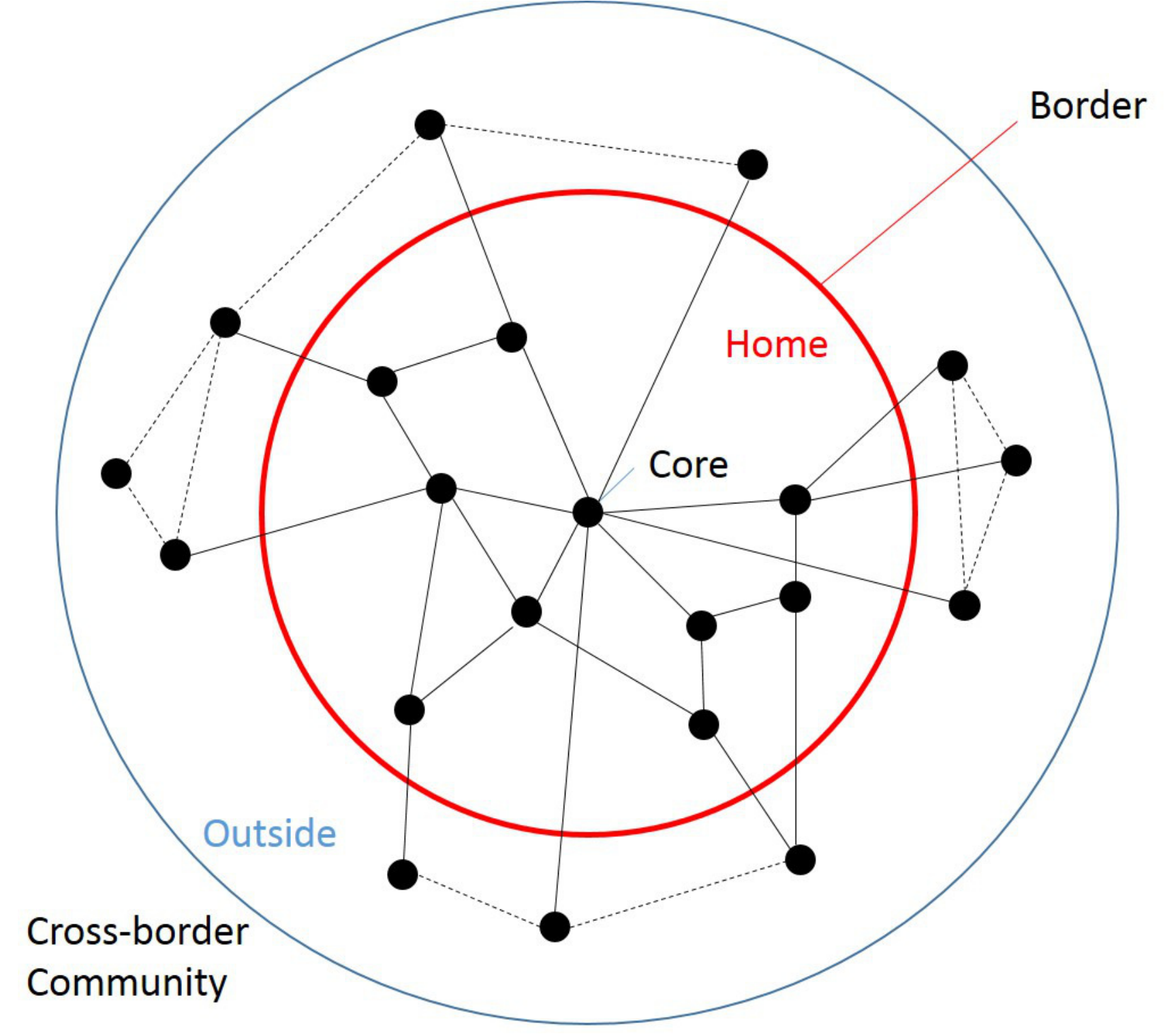}
 \end{center} 
 \caption{
 {\bf Outreach index.} The outreach index $O_s(\mathcal{N}_s)$ of a cross-border
 community measures the fraction of cross border and external ties (dashed lines
 in the plot), weighted by distance and relational intensity. The home base of the community is defined according to three criteria: : simple $|dQ|$, $|dQ|*S$, where $S$ is the node strengh, and internal link density $W_{int}$. When boundaries constrain the span of network communities -- as for European R\&D collaborations -- the outreach index lean towards zero. Conversely, if borders do not affect the shape of network communities -- like in the US --  $O_s(\mathcal{N}_s) \approx 1$, the topology of the network is conditioned by the presence of borders, which significantly reduces the probability 
 of cross-border connectivity.     
 \label{outreach}}
\end{figure}

The geographical dispersion is a pure spatial index and does not contain
any information about the network structure of the possible links
connecting the nodes embedded in space. Since this index is neither normalized nor weighted, it is inadequate for the comparison of different structures, 
like Europe and the US, 
where the distances are considerably different. 
Moreover, the geographic span does not measure how communities reach out. 
To do this, we introduce a new index, the \emph{outreach index}, defined as follows:

\begin{equation}
O_s(\mathcal{N}_s) = 1- \frac{\sum_{i,j \in \mathcal{N}_s} d_{ij}w_{ij}}{\sum_{i,j \in \mathcal{C}_s} d_{ij}w_{ij}}, \; O_s(\mathcal{N}_s) \in [0,1]
\end{equation}

\noindent where $\mathcal{N}_s$ is the home base of the community \emph{s} 
, $d_{ij}$ and $w_{ij}$ are
the distance and the weight of the link between nodes \emph{i} and \emph{j} and $\mathcal{C}_s$ is the community \emph{s} as before. 
The outreach index is defined as the ratio of all the weighted links except for the ones between the nodes \emph{i} and \emph{j} 
which are internal to $\mathcal{N}_s$, but still belonging to $\mathcal{C}_s$, with respect to the same quantity calculated 
for all pairs of nodes \emph{i} and \emph{j} belonging to $\mathcal{C}_s$.
Figure \ref{outreach} provides a schematic representation of the way in which the outreach index is obtained.

Multiple criteria can be used to select the home base $\mathcal{N}_s$:
\begin{enumerate}
 \item the home base is located in the region with the highest $|dQ|$. The regions with the highest $|dQ|$ can be defined as the core of the community, based
 on the intensity of intra-community ties. 
  \item the center of the community can be chosen as the one with the highest $|dQ|*S$, where $S$ is the sum of the weights of all outgoing and incoming links of a node. 
 This index accounts for both the role the node plays in 
 the intra-community connectivity ($|dQ|$) and the overall centrality of the region, as measured by the node strength ($S$).
 \item the area that scores the highest internal link density is selected as the home base. Intuitively, this criterion identifies the region with the highest share 
 of inner linkages in the community. In such a case the selected regions will be the biggest ones, regardless of where the core region is located.
\end{enumerate}

In our analysis we find that the above listed criteria tend to provide similar results. Therefore the choice betwenn them depends on the selected community detection method and data availability. 

\section*{Data} \label{sec:data}

The data analyzed in this study are drawn from the June 2012 release of the OECD REGPAT database \cite{webb2005,maraut2008} , which contains $~2.4\times10^6$ patent 
applications filed with the European Patent Office (EPO) from 1960 to the present.
In this database the geographical location of each patent inventor and applicant has been matched to one of the appropriate 5,552  regions in one of the
50 OECD or OECD-partner countries. This allows us to construct the geographical networks of patent co-inventorship. 

Starting from these data we define $w_{ij}$ as the number of links between regions $i$ and $j$.
In our network $w_{ij}$ will be equal to the number of patents
jointly invented by the two regions. We use a full-counting approach so that a patent with $N$($>1$) inventors accounts for $\sum_{i=1}^{N-1}(N-i)$ regional links (hence, patents with only one inventor do not appear in this network by construction). Therefore, we analyze a weighted undirected network of scientific and technological 
collaborations across regions. In the co-inventor network the intensity of a link between two regions is equal to the number of patents jointly invented by inventors located 
in those regions.

Patent data has long been analyzed to measure innovation outcome, just as patent co-inventorship has been used to study the network of innovators within and across national borders.
Recently, it has been found that scientific collaborations in Europe are much more constrained by spatial interaction than in the US \cite{crescenzi2007,andersson2009,chessa2013}. 
The European Union clearly represents a real case of transnational network since borders in this case are not only geographical but also political, administrative and 
cultural (states in the European Union differ by government, legislation, language and even religion). Conversely, state borders in the United States are of a different nature: 
despite being under the federal system the United States still share  the same central government, the same language and more or less the same culture. 
Thus we use the US innovation system as a benchmark to estimate the impact of national borders on European network formation.

In order to unveil these differences we compare the European and  US co-inventorship networks. Nodes are the NUT3 regions for Europe and the FIPS (Federal Information 
Processing Standard) geographical units for the USA, which corresponds to counties. (The Nomenclature of Units for Territorial Statistics (NUTS) is a geo-code standard for referencing the subdivisions of 
countries for statistical purposes. The nomenclature has been introduced by the European Union, for its member states. The OECD provides an extended version of 
NUTS3 for its non-EU member and partner states). Since we are interested in long-range connectivity across borders, only interactions that 
took place between different NUTS3 (or FIPS) are taken into consideration. That is to say, we do not consider self-loops in the following analysis. Nevertheless, our approach still
naturally extended to the case of directed weighted networks with self-loops.

\section*{Results} \label{sec:results}

Table \ref{table_1} reports the value of the \emph{geographical span} on our data.
Larger values of $D$ means that the members of the community are geographically spread out; at first glance one could 
conclude that US community members are more spread out than the European ones on average.  

\begin{table}
\begin{center}
\caption{European versus US communities: geographical span, $D_s$, and outreach index, $O_s(\mathcal{N}_s)$}
	\resizebox{8.7cm}{!}{
    \begin{tabular}{ |l|l|l|l|l|l|l|l|}
    \hline
   	 \multicolumn{4}{|c|}{Europe} & \multicolumn{4}{|c|}{United States} \\
  \hline
     Core & Country & $D_s$ & $O_s(\mathcal{N}_s)$ & Core & State & $D_s$ & $O_s(\mathcal{N}_s)$ \\ 
     \hline
	 Mannheim & DE & 1.9090 & 0.0509 & San Jose & CA & 13.8236 & 0.8775 \\ 
     \hline
	D\"{u}sseldorf & DE & 1.1016 & 0.0038 & New Brunswick & NJ & 7.8733 & 0.9379 \\ 
     \hline
	Paris & FR & 6.0087 & 0.0181 & Cambridge & MA & 11.2038 & 0.8911 \\ 
     \hline
	Berlin & DE & 2.2098 & 0.0477 & Cincinnati & OH & 3.9768 & 0.7525 \\ 
     \hline
	Stuttgart & DE & 1.4729 & 0.0087 & Philadelphia & PE & 9.6278 & 0.9792 \\ 
     \hline
	Eindhoven & NL & 1.5392 & 0.5440 & Minneapolis & MN & 7.9104 & 0.9803 \\ 
     \hline
	Munich & DE & 1.1793 & 0.0000 & Chicago & IL & 7.9630 & 0.9909 \\ 
     \hline
	Cambridge & UK & 2.5560 & 0.1335 & San Diego & CA & 17.8420 & 0.9288 \\ 
     \hline
	Helsinki & FI & 6.3203 & 0.6942 & Houston & TX & 6.9423 & 0.9803 \\ 
     \hline
	Nuremberg & DE & 1.5577 & 0.0223 & Cleveland & OH & 6.7129 & 0.9734 \\ 
     \hline
	Milan & IT & 3.4392 & 0.0309 & Raleigh & NC & 6.4801 & 0.9876 \\ 
     \hline
	Wien & AT & 1.9614 & 0.0537 & New Haven & CT & 11.3220 & 0.9641 \\ 
     \hline
	Barcelona & ES & 5.2386 & 0.5065 & Schenectady & NY & 6.7066 & 0.9718 \\ 
     \hline
	L\"{o}rrach & DE & 0.2957 & 0.6356 & Atlanta & GA & 5.7034 & 0.9364 \\ 
	\hline
	Average & EU & 2.6278 & 0.1964 & Average &  US & 8.8634 & 0.9394 \\  
	\hline
    \end{tabular}
    }
    \bigskip
    \caption*{The table compares cross-border communities in the US and Europe. Communities are identified based on the main city of the NUTS3 region with the highest $|dQ|$ (core). 
    We report the values of geographical span $D_s$ and the outreach index $O_s(\mathcal{N}_s)$ with respect to home country for European regions and states in the US.
    Home country has been selected base on the maximum $|dQ|*S$.}
  	 \label{table_1}
  \end{center}
  \end{table}

\subsection*{Distance distribution}

\begin{figure}[ht]
 \begin{center}
  \includegraphics[width=3in]{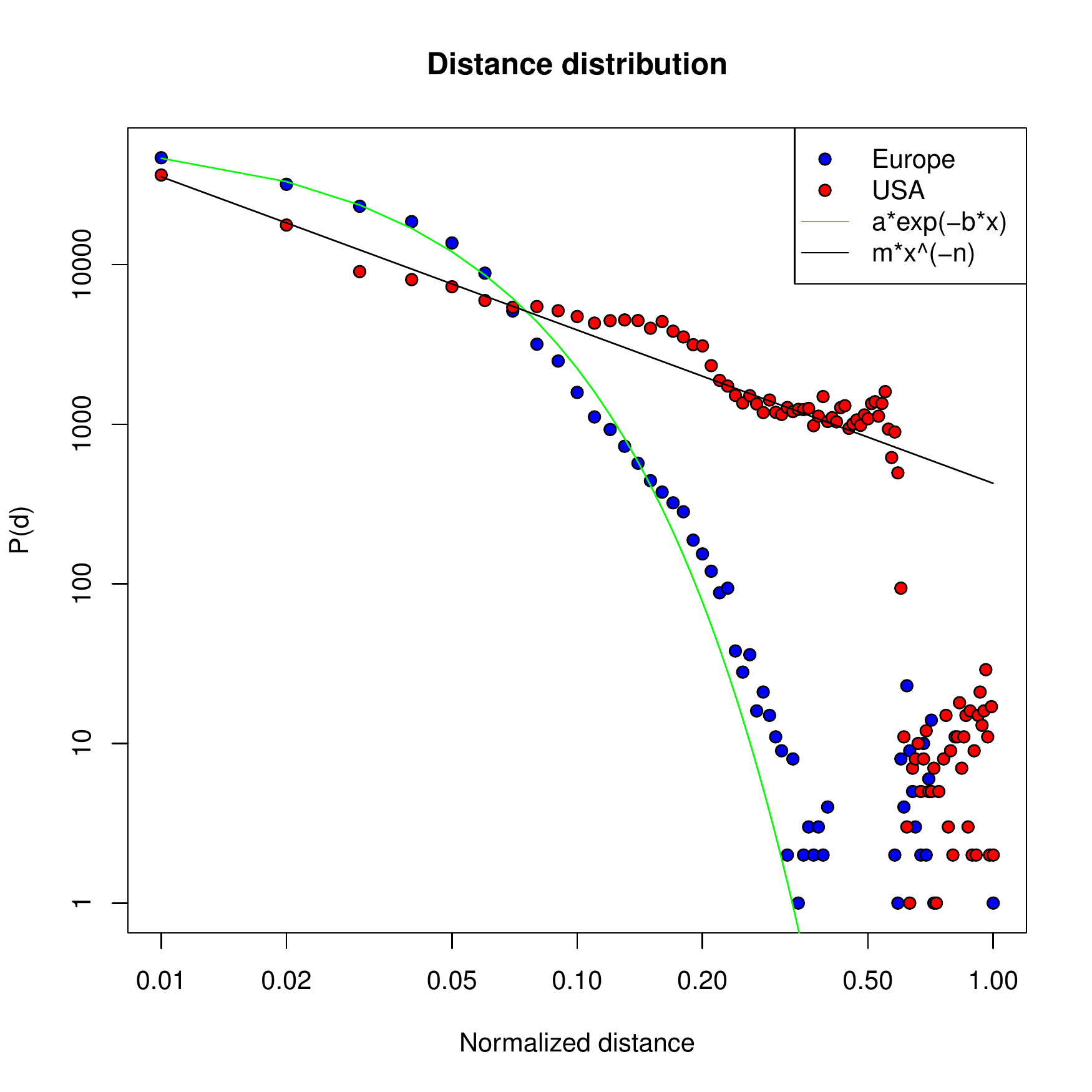}
 \end{center} 
 \caption{
 {\bf Distance distribution.} This figure shows the distance distribution of the links both for Europe (blue dots) and USA (red dots) in log-log scale and their best 
 fits: a power law $y=\beta d^{-\alpha}$ for the US (black), and exponential distribution $m \exp(-nx)$ for Europe (green), with $\beta=426.583\pm 29.474$, eps
 $\alpha=0.960\pm0.017$, $m=65030 \pm 570.5$, $n=33.7 \pm .35$.}
 \label{distance_distribution}
\end{figure}

It is well known that social interactions negatively depend on distance. More precisely, it has been shown that the probability of a tie between any pair of nodes decays with distance 
as a power-law $\sim d^{-\alpha}$ where $1 \leq \alpha \leq 2$ \cite{onnela2011,lambiotte2008}. Figure \ref{distance_distribution} compares the distance distribution of links in 
the European and 
US networks from 1986 to 2009. The two distributions clearly depict different behaviors: the US distribution is well approximated by a power-law as reported in the literature 
with an exponent $\alpha \approx 1$, whereas the European one shows an exponential behavior.

\begin{figure*}
\centering
\includegraphics[width=5in]{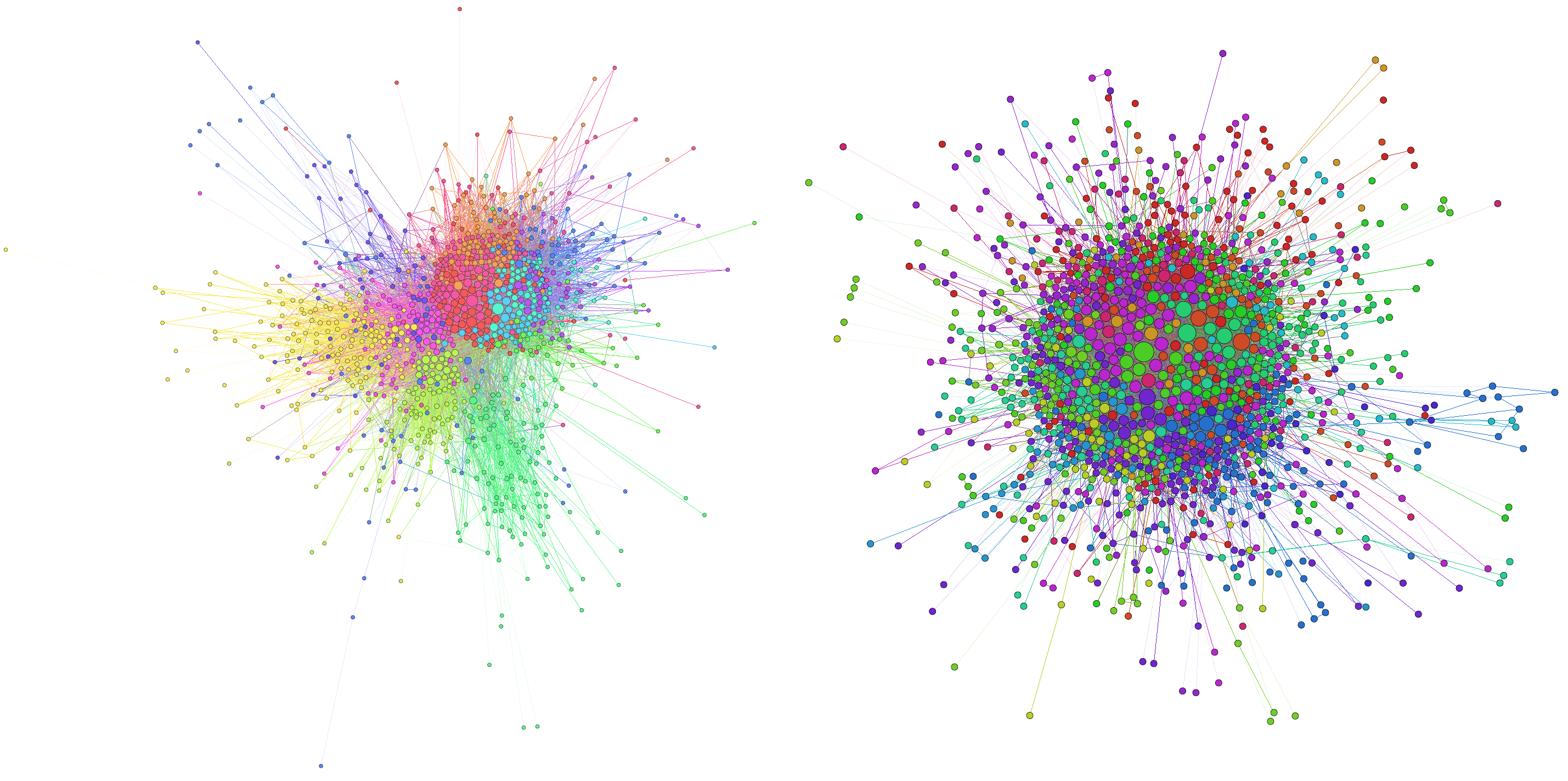} 
\includegraphics[width=5in]{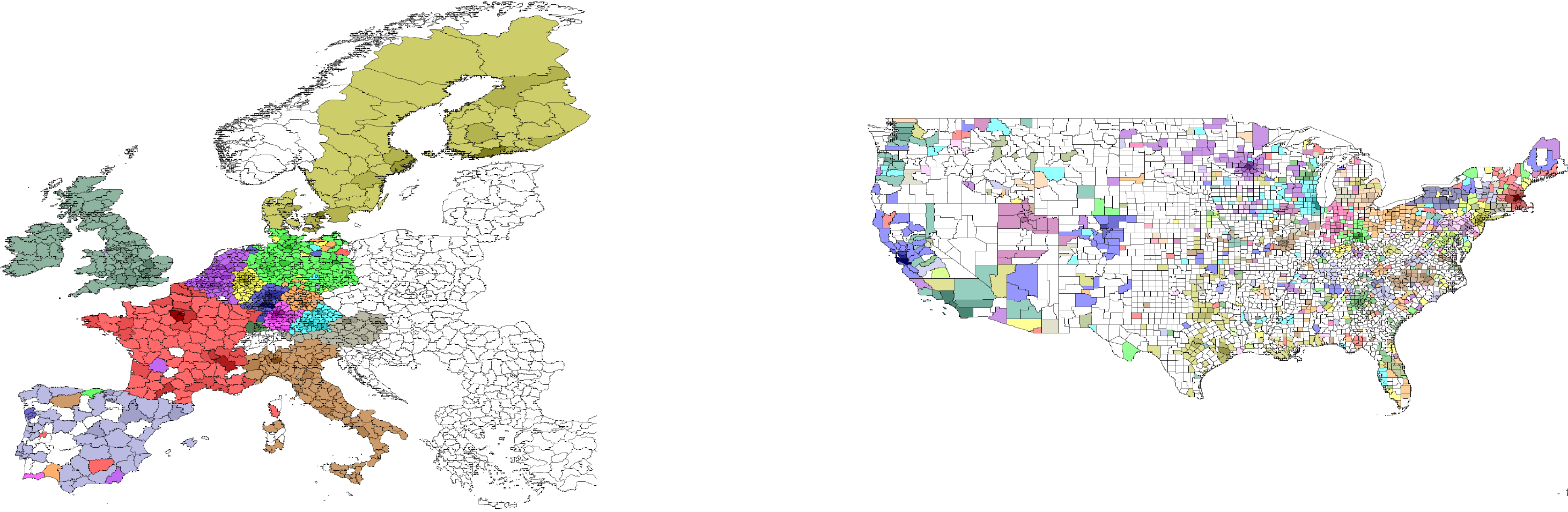} \\
 \caption{
 {\bf The community structure of the patent co-inventorship network: Europe versus the United States.} 
This figure shows the results obtained performing a community detection analysis both on the a) European and the 
b) US networks using the Newman-Girvan method. In this representation each community has been given a different color. 
It can be seen that, in the European case, the community structure almost perfectly matches the national boundaries 
of the 15 member states of the European Union. The only significant difference seems to be Germany, which is sectioned off 
into more than one community. The US community structure reveals almost an opposite behavior: communities are stretched  out
over more than one state and at great distances. Each community color is graduated according to the $dQ$ of the node: darker 
colored nodes have a higher $|dQ|$ so they are ``more central'', while lighter colored ones are less central since they have a lower
$|dQ|$.
The nodes with the highest $|dQ|$ are considered community core regions.
We use the same community color coding for networks (top) and maps (bottom). Maps and networks were generated using the open source software Gephi and QGIS, respectively.
    }
 \label{core_topological}
\end{figure*}

Figure \ref{core_topological} shows the results of the analysis performed using the
modularity method. For the sake of clarity, every node that corresponds to a geographical region, which is
geo-referenced and displayed on a map, is given the same color as the community
it belongs to. This results in nodes in different communities having different colors as well.

Figure \ref{core_topological} also shows the results of the core analysis
performed on the partition obtained using the Newman-Girvan Modularity.
In this representation each community has been given a different color. 
In the European case, the community structure almost perfectly matches the
national boundaries of the Member States of the European Union. 
The only significant difference seems to be Germany, which is sectioned off 
into multiple communities with an average size in the order of a Land
Region (NUTS2 level).
The US community structure reveals, on the other hand, a practically  opposite behavior:
communities are stretched out over more than one state, at great distances from
the alleged geographical core.
For each community, colors are graduated according to the $dQ$ of the node:
darker colored nodes have a higher $|dQ|$ and are therefore ``more central" while lighter colored ones 
are less central since they have a lower $|dQ|$.
We chose to define the nodes with the highest $|dQ|$ as the community core
regions.
Combined with our previous results regarding the different decay of connectivity at a distance (power-law in the US, exponential in Europe),
we clearly show that the presence of national borders in Europe has a strong role in shaping the topology of the network, both reducing
connectivity at a distance and constraining networks community in space.

As previously said, Europe is a genuine transnational network whereas the US system is not. Accordingly, their different behaviors do not come as a surprise. On the one hand, 
since the US innovation system is rather homogeneous, the probability for coast-to-coast interaction (up to a cut off distance of $\sim 10^3$ kilometers) is high. 
On the other hand, 
Europe is a collection of almost independent national systems of innovation (see Figure \ref{distance_distribution}). 
Namely, in the European case there is a cut off in the distribution due to strong country border effects, that eventually results in the exponential decay
behavior with a characteristic length that is roughly of the size of the average country diameter (about 363 Kilometers). The European network thus differs sharply from the US case, 
where the state border effect is almost negligible. In the US, scientific and technological communities span throughout the country 
without any characteristic scale. Moreover we find a power law decay of connectivity at a distance; even when we focus on a single European nation such as Germany, 
we still note some interesting differences with borders that play a stronger role in reducing connectivity between German L\"{a}nders when compared to their US counterparts.

Next we proceed to consider the outreach index of the communities. Before doing that, we must determine which one of the three criteria reported in the previous section is the most
appropriate. In Table \ref{table_3} we report the home base country of the communities we identify. As one can see, the outcome is the same in all the cases except for the ninth European community, for which we obtained
Denmark as the home base according to the first criterion, and Finland for the other two.
All in all, it turns out that the final result does not crucially depend on the method we use to identify the home country. Therefore, the outreach index has an high degree of universality.  
In the following analysis we will opt for the 
sensible solution of using a balanced method which takes both the topological centrality inside the community ($|dQ|$) and the total weight ($S$) attached to that node (second criterion) into account.
In the cases in which the community detection does not come from a modularity optimization and the $|dQ|$ value is not available, the third criterion can also be considered as a viable alternative.

\begin{table}
\begin{small}
\begin{center}
\caption{Home base of communities according to three different criteria}
	\resizebox{8.7cm}{!}{
    \begin{tabular}{|c|l|l|l|c|l|l|l|}
    \hline
   	 \multicolumn{4}{|c|}{Europe} &  \multicolumn{4}{|c|}{United States} \\
  \hline
 Community   &  $\mathcal{N}_s(dQ)$ & $\mathcal{N}_s(dQ*S)$ &$\mathcal{N}_s(W_{int})$ & Community & $\mathcal{N}_s(dQ)$ & $\mathcal{N}_s(dQ*S)$ & $\mathcal{N}_s(W_{int})$ \\ 
    \hline
	I & DE & DE & DE & I &  CA & CA & CA \\ 
    \hline
	II & DE & DE & DE & II &  NJ & NJ & NJ \\ 
    \hline
	III & FR & FR & FR & III &  MA & MA & MA \\ 
	\hline
	IV & DE & DE & DE & IV & OH & OH & OH \\ 
    \hline
	V & DE & DE & DE & V & PE & PE & PE \\ 
     \hline
	VI & NL & NL & NL & VI & MN & MN & MN \\ 
     \hline
	VII & DE & DE & DE & VII & IL & IL & IL \\ 
     \hline
	VIII & UK & UK & UK & VIII & CA & CA & CA \\ 
     \hline
	IX & DK & FI & FI & IX & TX & TX & TX \\ 
     \hline
	X & DE & DE & DE & X & OH & OH & OH \\ 
     \hline
	XI & IT & IT & IT & XI & NC & NC & NC \\ 
     \hline
	XII & AT & AT & AT & XII & CT & CT & CT \\ 
     \hline
	XIII & ES & ES & ES & XIII & NY & NY & NY \\ 
     \hline
    XIV & DE & DE & DE & XIV & GA & GA & GA \\ 
	\hline
    \end{tabular}
    }
    \bigskip
    \caption*{The table compares the home base of countries found for each community using three different criteria: simple $|dQ|$, $|dQ|*S$, where $S$ is the node strengh, 
    and internal link density $W_{int}$. Results do not vary significantly with 
    the only exception of the Nordic cluster that, when we use $|dQ|$, has its center in Denmark instead of Finland ($|dQ|*S$ and $W_{int}$).}
  	 \label{table_3}
  \end{center}
  \end{small}
  \end{table}

Table \ref{table_1} reports the value of the outreach index by choosing the home base according to the second criterion. 
Given that the \emph{outreach index} always lies between
0 and 1, we are allowed to compare the outreach of European communities with US counterparts (see Table \ref{table_1}).
 As expected, the outreach value is about 0 for almost every
community in Europe, while this value is always close to 1 in the US. This means that the
communities in the United States undertake more outreach than in Europe. However, we should remember
that the United States are not a truly transnational network, and accordingly it makes sense to compare the US with Germany as we did before. 
Indeed, community detection showed that Germany behaves differently and
splits into several sub clusters. Then, if we take the NUTS2 level L\"{a}nd (the German equivalent of a US state) as the reference nation $\mathcal{N}_s$ instead of Germany as a whole (which is NUTS1), the outreach values are sensibly different. They become comparable to the US values ranging from .49 for the region centered around Munich (Oberbayern) to .95 for the region of Mannheim (Karlsruhe).

\subsection*{Simulations} 

\begin{table*}[tbh!]
\begin{center}
\caption{Simulated Outreach index.}
	\resizebox{12cm}{!}{
    \begin{tabular}{ |l|l|l|l|l|l|l|c|c|c|}
    \hline
   	\multicolumn{1}{|c|}{} & \multicolumn{5}{|c|}{Outreach index, simulated} & \multicolumn{3}{|c|}{Parameters}\\
    \hline
     $\mathcal{N}_s$ & $d = 0$ &$d = 10^1$ & $d = 10^2$ & $d = 10^3$ & $d = 10^4$ & $\gamma_{in}$ & $\gamma_{out}$ & $\gamma_{across}$\\ 
     \hline
	FR & 0.019 $\pm$ .002 & 0.022 $\pm$ .003 & 0.034 $\pm$ .005 & 0.121 $\pm$ .007 & 0.285 $\pm$ .019 & 1 & 0 & 0.08\\ 
    \hline
	NL & 0.533 $\pm$ .021 & 0.542 $\pm$ .021 & 0.538 $\pm$ .025 & 0.544 $\pm$ .022 & 0.534 $\pm$ .017 & 1 & 0.1& 0.10\\ 
     \hline
	FI & 0.711 $\pm$ .018 & 0.721 $\pm$ .015 & 0.722 $\pm$ .018 & 0.721 $\pm$ .025 & 0.725 $\pm$ .011 & 1 & 0.5 & 0.20\\ 
     \hline
	CA & 0.887 $\pm$ .011 & 0.889 $\pm$ .009 & 0.884 $\pm$ .009 & 0.889 $\pm$ .008 & 0.886 $\pm$ .008 & 1 & 0.1 & 0.20\\ 
     \hline
    \end{tabular}
    }
    \bigskip
    \caption*{ TThe table reports the values of the simulated outreach index $O$
    obtained for the 4 cases FR, NL, FI, CA and for different values of the
    separating distance $d$. We use here different values of $\gamma$ to
    differentiate the four cases of the simulation: $\gamma_{in}$ is the network density within borders, 
    $\gamma_{out}$ is the network density outside borders and $\gamma_{across}$ regulates 
    cross-border links between the inner and the outer part of the community.}
  	 \label{table_2}
\end{center}
\end{table*}

The inspection of the outreach index for the US and Europe reveals the existence of four main community types: the French case, 
which exemplifies European national communities, the Benelux and Nordic clusters, which are two cases of regional
integration, and the California case, which is representative of the general behavior of
long-range out reaching communities in the United States. 
In order to uncover the internal mechanics involved in the creation of these
cases we reproduced the relevant patterns that emerged from real data 
by simulating the internal structure of each community.

The artificial model is defined as follows. Out of a total number $N$ of nodes
we decide what fraction of them, say $N_i$, to place into the central/home
region and what fraction $N_e$ to place into the the external one(s). For the
sake of simplicity we choose to shape the regions as circles whose radii are proportional to the number of nodes, so that 
the more the nodes the bigger the region.

As the regions belonging to the same community can either be adjacent to each other or not, so we introduce 
the parameter $d$ to regulate the spatial separation between them. 

Once the nodes have been placed into communities, we randomly create links
between them until the maximum number of links is reached.
In general, if $M$ is the total number of possible links $M=N(N-1)/2$ for an
undirected network, we determine the density of the network, $\gamma$, as a number between 0 and 1, so that the total
number of links will be  $P =\gamma M$.

We fine tuned different network densities according to the number of regions that belong to the community and the number of nodes that each one of them contains. 
Thus we have different $P$'s regulating internal links in the central and the
external regions, cross-border links between the external regions and the central one and, finally, 
links between external regions (if more than one).
  
\begin{figure}
\centering
\includegraphics[width=3.6in]{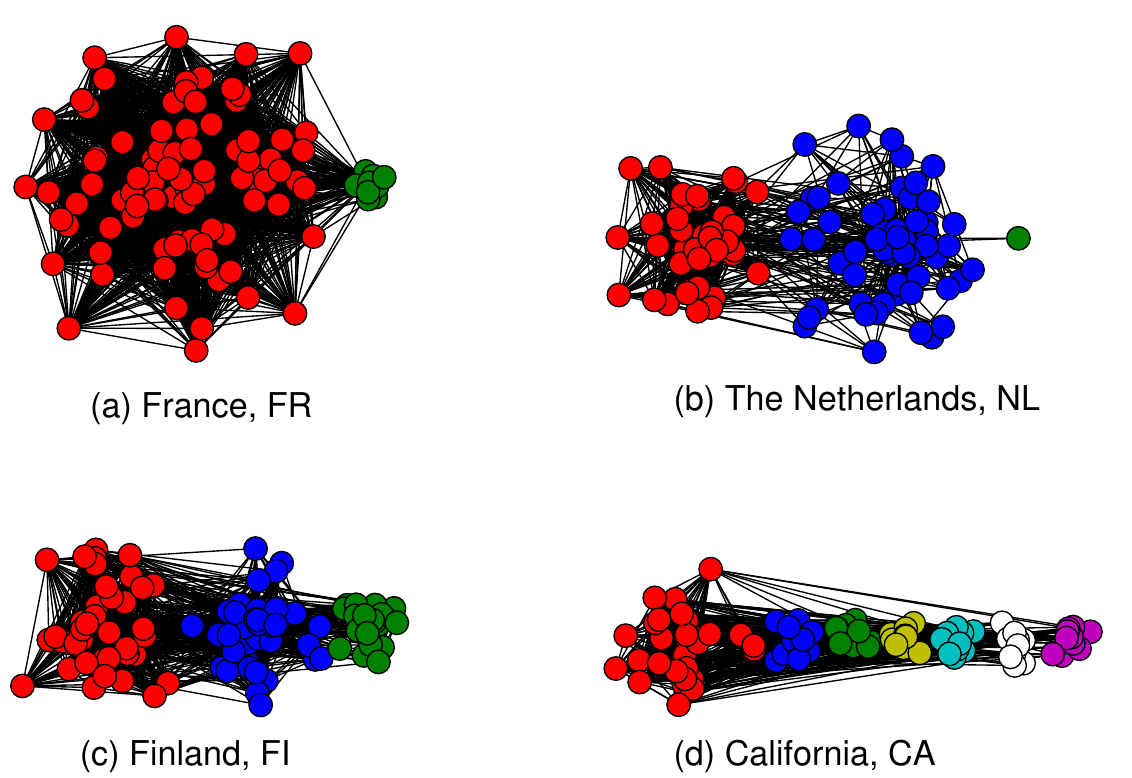} 
        \caption{ {\bf Simulations.} This figure shows the simulations reproducing the 4 different cases that we identified as France
         (FR), the Netherlands (NL), Finland (FI) and California (pdfCA). 
         In the FR case there is a main community, which is very well connected in the
         inside ($\gamma\sim 1$), with few links that reach out to external
         regions. The NL and FI cases are intermediate with well-structured external
         regions that still interact with the central one. In the last
         case, CA, there is a strong central region with many, and progressively
         distant, small regions. The FI and CA cases present similar outreach
         indices due to the fact that, by definition, a huge mass of links in
         the immediately external regions is equivalent to having just a few interconnected nodes
         at great distance.}
  		\label{fig:simulations}
\end{figure}

We then calculated the outreach index for different values of $d$ (see Table \ref{table_2}) for these networks. 
Even when $d=0$, the case for which there is no separation among the
regions (see Figure \ref{fig:simulations}), in addition to a set of parameters
extracted from the data, the model closely reproduces the spatial organization
of the four real cases. As we can also observe in Table \ref{table_2}, the
simulated outreach indexes are similar to the empirically observed values.

\section*{Concluding Discussion} \label{sec:discussion}

The role of distance in spatially embedded complex networks has been recently investigated. Empirically speaking, it has been found that connectivity tends to decay with 
distance according to a power-law relationship \cite{onnela2011}. This is in line with previous results in the economic literature, where an inverse power relationship has been 
repeatedly observed 
in gravity-like models of international trade, human migration and foreign investment \cite{disdier08}. Despite this growing body of evidence regarding complex networks 
in space, still little is known about the role of national borders in the formation of cross-national networks. In this paper we aim at understanding more about this role by analyzing the structure of 
network 
communities within and across borders. We show that, while the connectivity of US scientific communities decays as a power of distance, European scientific communities 
tend to be confined within national borders. We introduce a new measure for the outreach of network communities across borders and confirm our results via simulations. All in all, 
our findings reveal that Europe is still a collection of national systems of innovation and the European Research Area is still far from becoming reality \cite{chessa2013}. 
Our methodological approach can be used to keep track of the progress toward
the integration of the European Research Area. More in general, the outreach index we discuss in this paper is worth using to detect the impact of borders on the 
formation and dynamic evolution of spatially embedded networks.

\section*{Acknowledgments}
Authors acknowledge funding from the PNR project “CRISIS Lab”. MR acknowledges funding from the MIUR (PRIN project 2009Z3E2BF) and FWO (G073013N). Federica Cerina 
gratefully acknowledges Sardinia Regional Government for the financial support of her PhD scholarship (P.O.R. Sardegna F.S.E. Operational Programme of the Autonomous 
Region of Sardinia, European Social Fund 2007-2013 - Axis IV Human Resources, Objective l.3, Line of Activity l.3.1.).

\bibliographystyle{unsrt}
\bibliography{biblio}

\end{document}